\documentclass[12pt]{article}
\usepackage{graphicx}
\usepackage{color}
\usepackage{longtable}
\usepackage{hyperref}
\usepackage{latexsym}
\usepackage{amsmath}
\usepackage{amssymb}
\usepackage{dsfont}
\usepackage{verbatim}
\usepackage{bm}
\newcommand{\I}{\mathrm{i}}  
\newcommand{\FC}{\;,}
\newcommand{\FD}{\;.}

\newcommand{\dbar}{\overline{d}}
\newcommand{\ot}{\ensuremath{\frac{1}{2}}}
\newcommand{\Npi}{\ensuremath{{N\pi}}}
\newcommand{\unitmatrix}{\mathds{1}}
\newcommand{\myparagraph}[1]{~\vspace{1mm}\\ \noindent{\bf #1}}
\newcommand{\refcite}[1]{\cite{#1}}
\begin{document}
\thispagestyle{empty}

\date{\today}


\begin{center}
\normalsize PION NUCLEON SCATTERING: \\SOME RESULTS FROM LATTICE 
QCD\footnote{Contribution to the Nstar 2013 Workshop, Peniscola, Spain, May 27-30, 2013}
\end{center}

\begin{center}
\normalsize C. B. LANG\footnote{\small Speaker at the conference}$^{,+}$~~and V. VERDUCI$^{+,\#}$
\vspace{5mm}\\ \small
$^+$Institut f\"ur Physik,  Universit\"at Graz, A--8010 Graz, Austria\\ \small
$^\#$School of Physics and Astronomy, University of Edinburgh,\\  \small Edinburgh EH9 3JZ, Scotland
\end{center}


\begin{abstract}
Including the meson-baryon (5 quark) intermediate states in a lattice simulation
is challenging. However, it is important in order to obtain the correct energy
eigenstates and to relate them to scattering phase shifts. 
Recent results for the negative parity nucleon channel and the problem
of baryonic resonances in lattice calculations are discussed.
\end{abstract}


\section{Motivation}
The Euclidean space-time lattice regularisation is the only controllable non-perturbative regularisation of QCD. It defines the quantum field theory as a limit from  a finite to an  infinite number of points and from  non-zero to vanishing lattice spacing. At any step in the limit one can
in principle solve the functional integrals and determine values for $n$-point amplitudes by numeric means. The last decades have
shown an enormous progress towards computing hadronic properties that way, mainly by Monte Carlo methods on very large
computers.

Due to the finiteness of the spatial volume, correlation functions of the type $\langle \mathcal{O}(t)\mathcal{O}(0)\rangle$ have a discrete energy spectrum -- as opposed to the continuous spectral function in the continuum. The ground states can be computed from the correlation function using hadronic operators in the corresponding quantum channel.

In Nature the hadrons are, however, stable (w.r.t. hadronic decays) only in a few cases (proton, pion, kaon,...). In the lattice world, where the quark mass may be set to unphysically large values and where the
momenta are quantized due to the finite spatial volume, some decay thresholds lie higher and the respective lowest state appears to be (artificially) stable. 
On the other hand, two- (and more-) hadron intermediate states should have an impact on the energy spectrum. Most of those have not been observed up to now.
In recent years there have been several studies determining baryonic excitations.\cite{Edwards:2011jj,Mahbub:2012ri,Engel:2013ig,Alexandrou:2013fsu}\footnote{For brevity we omit the references to the equally important detailed studies on meson excitations.} One common feature was the absence of meson-baryon decay states (except for possibly the $s$-wave states, see below). This was especially blatant in the $\rho\to\rho$ studies, which on the lattice should exhibit $p$-wave
$\pi\pi$ intermediate states. These should show up (but did not) as discrete energy levels of the spectral decomposition of the
hadron correlation function. In the meson sector the cure was to include meson-meson interpolators in the set of
operators of the correlation matrix. Once these were allowed for, the diagonalization of the
cross-correlation matrix exhibited the levels that had been missing before. The earlier observed levels, that had been interpreted as
resonance excitations appeared shifted, depending on resonance width and lattice volume.

In the mentioned baryon studies, the baryon interpolators were exclusively three-quark operators and the statements above applied. Possible exceptions may be $s$-wave states like the negative parity nucleon $N(\ot^-)$ where recent results indicated
a coupling the $N\pi$ even when using only three-quarks correlators.\cite{Mahbub:2012ri,Engel:2013ig,Alexandrou:2013fsu}

In order to clarify the situation we studied the negative parity nucleon channel including  $s$-wave $N\pi$ (4+1 quark) operators.\cite{Lang:2012db} Indeed we find significant differences in the energy spectrum, as will be discussed below.
Our study is for gauge configurations with two mass degenerate dynamical quarks, generously provided by 
the authors of Ref. \refcite{Hasenfratz}. The used parameters correspond to a lattice spacing of 0.1239 fm on a $16^3\times 32$ lattice with a pion mass of 266 MeV. Details can be found in Ref.~\refcite{Lang:2012db}. \vspace{-2mm}

\section{Tools}
\myparagraph{Operators:} For determining the low lying energy levels we applied the so-called variational analysis. One computes the cross-correlation matrix between a sufficiently complete set of hadronic lattice operators with the chosen quantum numbers ($J^P=\ot^-$ in our case). For the nucleon interpolator we use the standard operator (on a given time slice)
\begin{equation}\label{eq:defN}\vspace*{-2mm}
(N_\pm^{(i)})_\mu(\vec p=0)=
\sum_{\vec x}\epsilon_{abc}\, \left(P_\pm\,\Gamma_1^{(i)}\, u_a(\vec x)\right)_\mu\, 
\left( u_b^T(\vec x)\, \Gamma_2^{(i)}\, d_c(\vec x)\right)\FD
\end{equation}
$(\Gamma_1, \Gamma_2)$ can
assume the three values $(\unitmatrix,C\gamma_5)$, $(\gamma_5,C)$ and
$(\I\unitmatrix,C\gamma_t\gamma_5)$ for $i=1,2,3$. $C$ denotes the charge
conjugation matrix, $\gamma_t$ the Dirac matrix in time direction, and
$P_\pm=\ot(1\pm\gamma_t)$ the parity projection. We sum over all points of the
time slice in order to project to zero momentum.The pion interpolator is, e.g., 
$\pi^+(\vec p=0)=\sum_{\vec x} \dbar_a(\vec x) \gamma_5 u_a(\vec x)$. Summation over the color index $a$ is implied.

\begin{figure}[t]
\begin{center}
\includegraphics[height=4.5cm,clip]{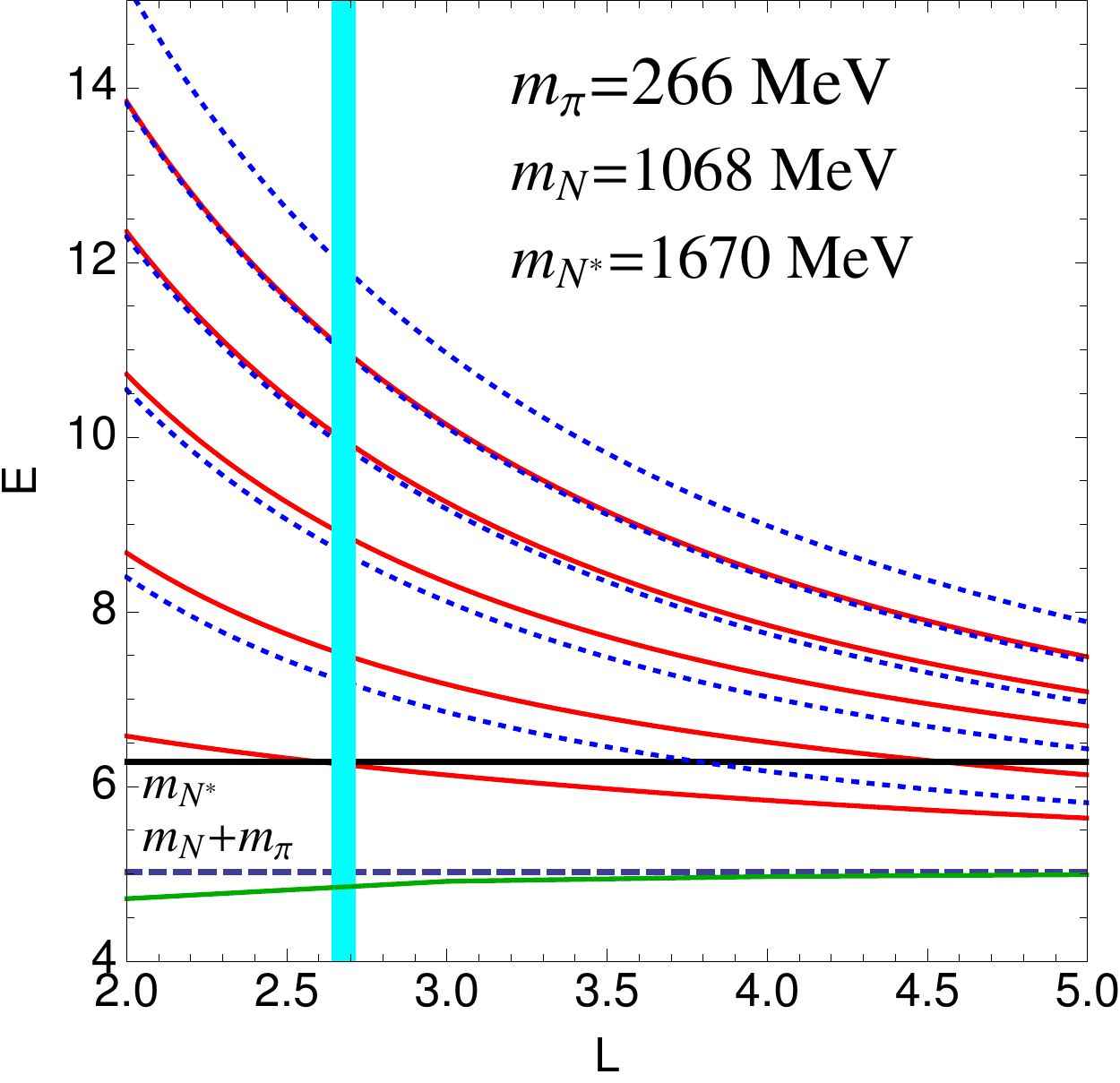}\hfill
\includegraphics[height=4.5cm,clip]{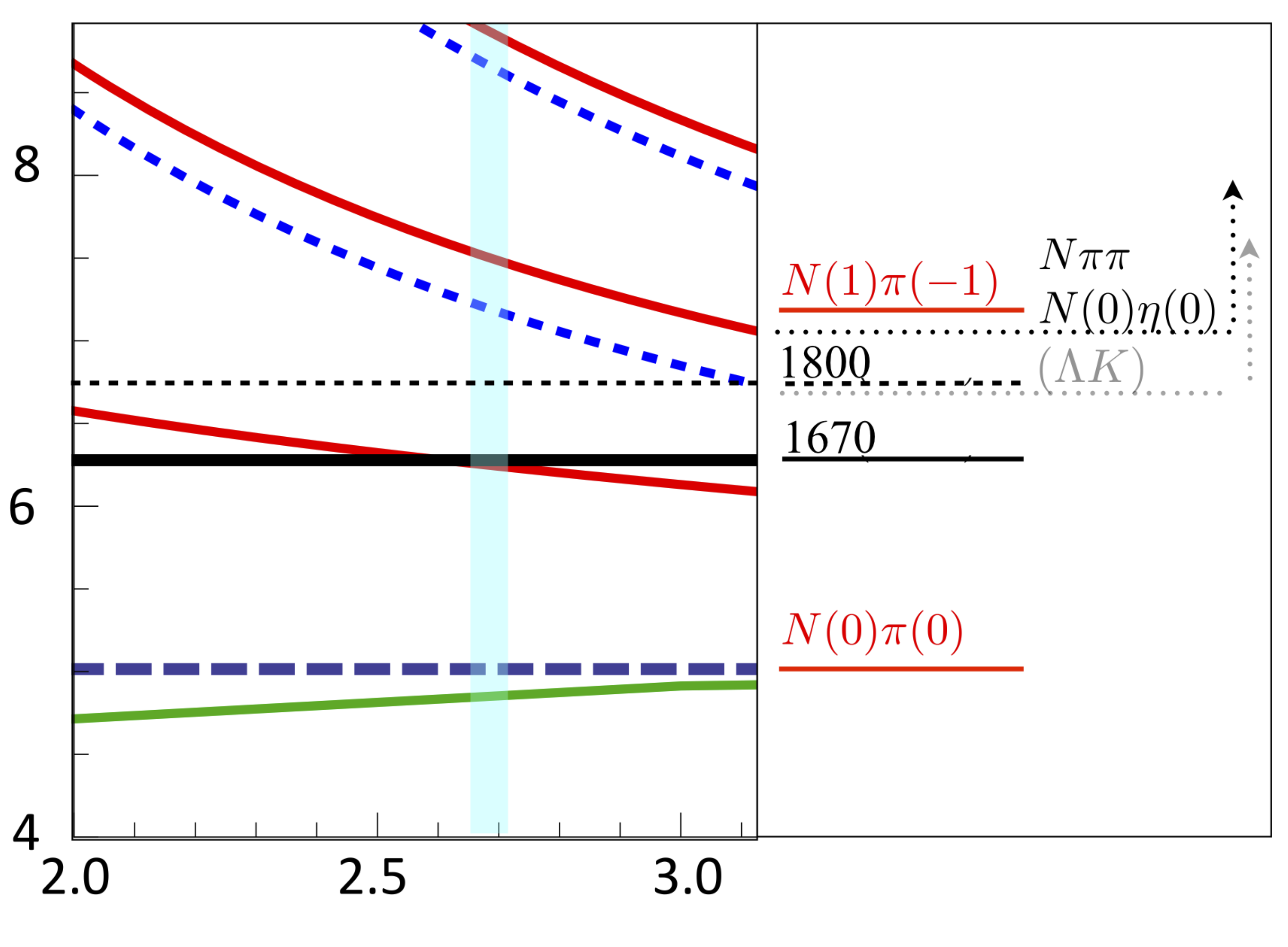}\hfill
\includegraphics[height=4.5cm,clip]{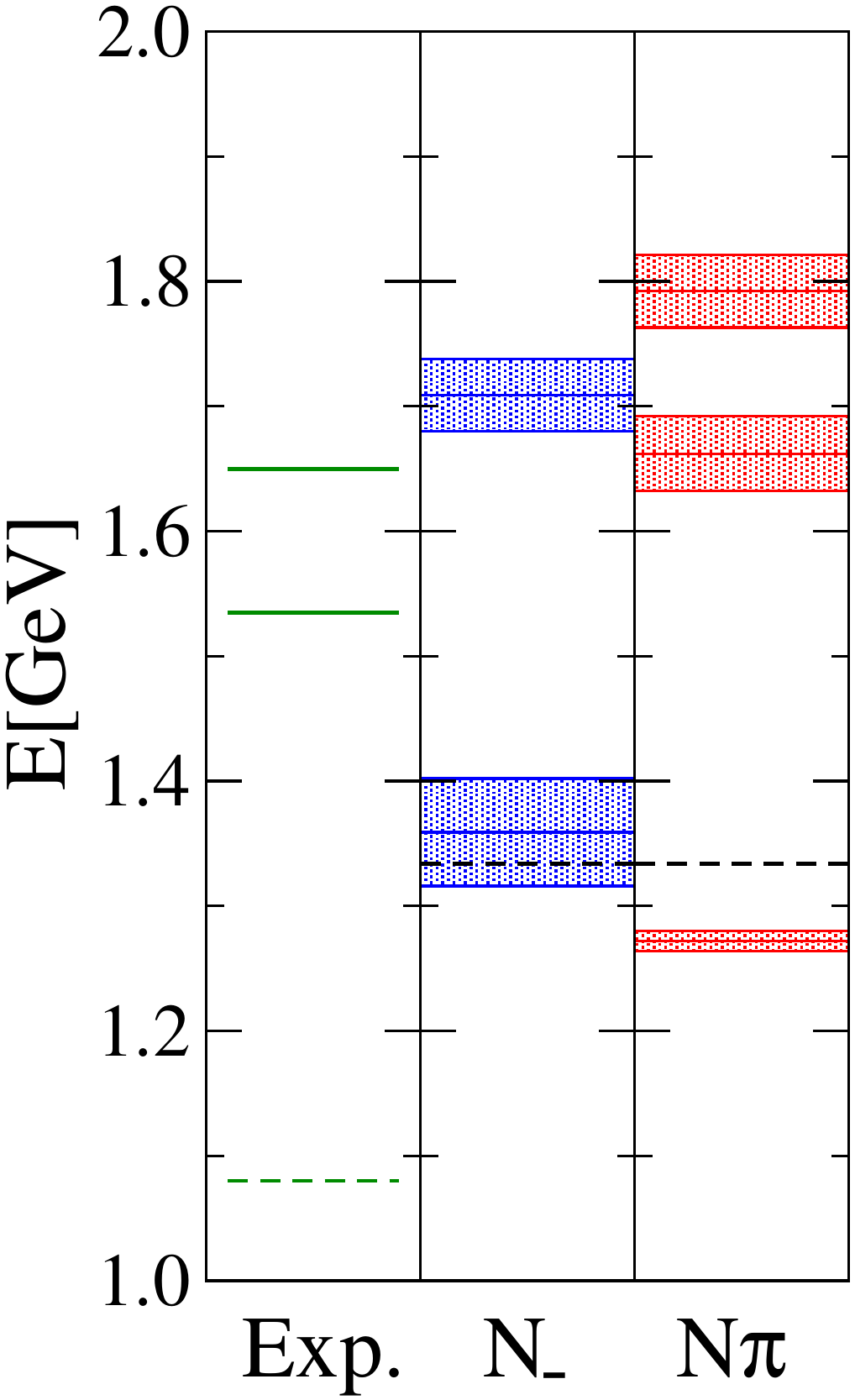}
\end{center}
\caption{The $N\pi$ levels without (blue) and with (red and green) interaction assuming a Breit-Wigner shape phase shift
for our situation. All energies as well as the spatial size $L$ use units of $m_\pi=266$ MeV. The lowest $N\pi$ level approaches the threshold  from below towards $L\to\infty$.  Middle: Details, possible coupled channels thresholds are indicated; note that the
numbers 1670 and 1800 correspond to the $N^*$ resonances in a world with $m_\pi=266$ MeV, thus they come out higher than their physical values. R.h.s.: Comparison of the energy levels observed without (middle) and with (right) including $N\pi$ operators.}\label{fig:lug}
\end{figure}
For the $\Npi$ system in the rest frame the leading $s$-wave
contribution comes from the interpolator with both particles at rest,
$\Npi(\vec p=0)=\gamma_5 N_+(\vec p=0)\pi(\vec p=0)\FC
$
where $N_+$ denotes the positive parity nucleon and the factor $\gamma_5$ 
ensures negative parity for the interpolator. We project to isospin $\ot$.
In the distillation approach  we
choose Å$N_v=32$ and $64$ for each of the 3-quark nucleon interpolators and $N_v=32$  for the $\Npi$ channel. We then select the most suitable combination and diagonalize a $7 \times 7$ correlation matrix to obtain the results shown in Fig. 1.
\myparagraph{Variational analysis:}
We determine the energy levels of the coupled $N$
and  $N\,\pi$ system with help of the variational 
method.\cite{VarMeth} For a given
quantum channel one measures the Euclidean cross-correlation matrix $C(t)$
between several interpolators living on the corresponding Euclidean time slices. The
generalized eigenvalue problem  disentangles the
eigenstates $|n\rangle$ and from the exponential decay of the
eigenvalues
one determines the energy values of the eigenstates by exponential fits to the
asymptotic behavior. 
\myparagraph{Wick contractions:}  In order to compute this correlation matrix we had to first compute the Wick decomposition of the correlators in terms of the quark propagators. For $N\to N$ these are of two types, while the complete 
$(N_-, N_+\pi)\leftrightarrow(N_-, N_+\pi)$ system requires the evaluation of 29 graphs.\cite{Lang:2012db}
\myparagraph{Distillation:} Among all these contraction terms, there are some  involving backtracking propagators where special tools are necessary to obtain statistically reliable signals. We used the so-called distillation method.\cite{Peardon:2009gh} On a given time slice one
introduces separable quark smearing sources derived from the
eigenvectors of the spatial lattice Laplacian. This allows high flexibility due to the disentanglement of the computation of the quark propagators (``perambulators'') and the hadron operators.
\myparagraph{Phase shifts:}  L\"uscher\cite{Luscher} derived a relation between the energy levels at finite volume and the phase shifts of the infinite volume, valid in the elastic region. For the pion mass used, the nucleon mass lies roughly 130 MeV above the physical one. Similar 
mass shifts occur for other resonances. We argue (see Fig.~\ref{fig:lug}), that the thresholds for $N\eta$, $N\pi\pi$ and $\Lambda K$ lie (closely) above the
two lowest states ($N(1535)$ and $N(1650)$, shifted upwards as well), i.e., above 1800 MeV. Also the $N(1)\pi(-1)$ state (the value ``1'' refers to the smallest momentum unit $2\pi/16a$ for our spatial lattice extent) lies higher. We did not include these meson-baryon operators in our
set. Thus we apply  L\"uscher's relation up to the region of the second resonance. More detailed studies would have to include more operators and deal with the coupled channel problem. For a discussion of pole shifting due to changes in the mass see Ref. \refcite{Doring:2013glu}.\vspace{-2mm}

\section{Results}

Fig. \ref{fig:lug} (r.h.s.) demonstrates that the energy spectrum changes when allowing for the $N\pi$ coupled channel. We find now a
clear signal for the lowest $N(0)\pi(0)$ state, lying closely below threshold, as expected. Indeed also the quality of the energy levels improves. In a Breit-Wigner fit to the
corresponding phase shift values the resonance masses lie approximately 150 MeV above the physical values, similar to the nucleon mass, due to the unphysical pion mass of 266 MeV. Further details can be found in Refs.~\refcite{Lang:2012db} and \refcite{Lang:2013eca}.\vspace{-2mm}
 
\section*{Acknowledgements}
We thank Michael D\"oring, Georg Engel, Christof Gattringer, Leonid Glozman, Meinulf G\"ockeler, Daniel Mohler, Colin Morningstar, Sasa Prelovsek and Akaki Rusetsky for many discussions. Special thanks to Anna Hasenfratz for providing the dynamical configurations and to Daniel Mohler and Sasa Prelovsek for allowing us to use the perambulators derived in another project. V.V. has been supported by the Austrian Science Fund (FWF) under Grant No. DK W1203-N16.

\end{document}